# COLLABORATIVE DIGITAL LIBRARY OF HISTORICAL RESOURCES: EVALUATION OF FIRST USERS

**Abrizah Abdullah and A.N. Zainab**
Department of Information Science
Faculty of Computer Science & Information Technology
University of Malaya, 50603 Kuala Lumpur, Malaysia
e-mail: abrizah@um.edu.my; zainab@um.edu.my

*ABSTRACT*
*This paper describes the digital library of historical resources, a research project which involves building a testbed for the purpose of developing and testing new collaborative digital library functionality and presents an initial analysis of the digital library's public use on the web. The digital library is modeled to focus on serving secondary students information needs in conducting history projects. As such, in the implementation of the digital library, the use of online resources would be an integral part of history project-based learning activities. Students should be enabled to access digital resources, create and publish their own documents in the digital library and share them with others. As a testbed system, the collaborative digital library known as CoreDev has demonstrated its capabilities in serving an educational community as has been reflected by the positive feedback on the functional requirements from 44 users. Over 75% of the respondents in the user survey considered themselves capable of using the digital library easily. The beta tester demographics (n = 105) indicate that the digital library is reaching its target communities.*

**Keywords:** Collaborative digital libraries; User testing and evaluation; Digital libraries in education; Usability study; Historical resources; School projects

## INTRODUCTION

Digital library for education grew from grassroot efforts of teachers, students and researchers working collaboratively to create a library of educational resources and services to support K-12 teaching and learning. Massulo and Mack (1996) succinctly summarise the three roles digital library can play in K-12 education: as a resource for teaching in curriculum development; as a resource for learning to enrich students' experience; and as an authoring space in support of students learning. Digital libraries not only offer innovative strategies for learning opportunities, but they also can make a significant impact on enhancing and improving ICT and information literacy skills among students and teachers because hosting of information, retrieving and handling information from the Internet requires a fair amount of computer skills and Internet literacy (Zainab, Abdullah and Anuar, 2003) as well as information literacy Abdullah, Zainab and Yu, 2006)

Research shows that Internet technologies, such as digital libraries, have been of tremendous use to students' project works. Blumenfeld et al. (1991), Grant (2002), and Sidman-Taveau and Milner-Bolotin (2004) found that project-based learning is especially effective in enhancing student motivation and fostering higher order thinking skills, especially when supported by Internet technology. Durrance and Fisher (2003) indicated that the ability to use Internet technology as a tool is very significant in helping students to support their project work. Lynch (2003) contends that the project or resource-based learning movement has given rise to considerable interest in the use of online information resources as the basis for student-centred learning.

The approach to use digital libraries to publish and share resources to support project-based learning in the Malaysian educational context is no doubt forward-looking. The purpose of the digital library in this research is to provide the learning community with an experience in collaboratively building a digital library of history project reports, which indirectly allow members of the community to be aware and be actively involved in e-publishing as well as enhances member's ICT literacy skill. The digital library would benefit both its direct stakeholders – students who would be the creator and publisher of digital history project works, and teachers who would be given the experience of managing digital information. The collaborative resource development digital library in this study is a community-owned and governed digital library offering easy access to electronic resources on Malaysian history at all secondary educational levels. The resources are designed to support students' research in the form of project-based learning. Such resources, which should be evaluated by History subject teachers, include project reports, historical texts, images, audio, video, and links to relevant websites. Collaborators such as students and teachers, educational or historical institutions maintain local storage of these resources on their own servers, which are then accessed via the database of searchable metadata records that describe these resources. The users include learners and teachers in all venues, many of whom are also resource contributors, who develop educational materials, provide historical knowledge, and evaluate the digital library's holdings. These students and teachers will be partners in digital resource development as content developers and content managers respectively, and it is these partners who will form the nucleus of the collaboration. To date, the collaborative digital library, named CoreDev (http://coredev.fsktm.um.edu.my), has developed community structures, a strategic plan, and a useful collection of about 700 resources of various types and format.

**A BRIEF REVIEW OF LITERATURE**

Studies on use and usability of digital libraries explore users' receptivity in order to determine usage for a long time to come. Prior work on students use of digital libraries has concluded that they encounter barriers to effective information retrieval; not knowing what information is needed; not knowing where to find the information once it is known it is needed; not knowing sources of information exist; finding that no source of information exists; inaccurate or inappropriate information retrieved; and delays encountered in information retrieval (Abdullah, 2007b). Thong et al. (2004) identified three categories of external factors leading to a greater user acceptance of digital libraries: interface characteristics, organisational context, and individual differences. Interfaces characteristics include terminology clarity, screen design and navigation clarity, while organisational context pertains to the system relevance, system accessibility and system visibility. Individual differences include system efficacy, computer experience and domain knowledge. The researchers also recommended ways to increase user acceptance and believe that with the recommendations, digital libraries will be able to entice more users to discover and be adopted. They further emphasised that organisational context is critical to user acceptance of digital libraries. As such, digital libraries must be visible to users; users must be aware of the benefits of using digital libraries and their existence. Thong's research supports the mere exposure effect (Zajonc, 1968) where exposure to digital libraries can change users' attitude for the better. They recommended that the existence of a digital library be publicized, and orientation programmes be introduced to promote the digital library among potential users.

A number of studies examined the usability of digital libraries. It is through usability testing that researchers have started to address the role of the user in system design. In its most basic formulation, usability has been defined as "*a system's capability in human functional terms to be used easily and effectively by the specified range of users, given specified training and support, to fulfil a specified range of tasks, within a specified range of environmental scenarios*" (Shakel quoted in Dillon, 1994). Nielsen (2003) considers that the usability of a system can have five quality components namely learnability (how easy is it for the users to accomplish basic tasks the first time they encounter the design), efficiency (how quickly can they perform tasks once users have learned the design), memorability (how easily can they reestablish proficiency when users return to the design after a period not using it), errors (how many errors do users make, how severe are these errors, and how easily can they recover from the errors) and satisfaction (how pleasant is it to use the design). He considers that a usability test can be trustful enough with five users and indicates that by testing a system with five users it is possible to identify a great part of the usability problems (about 85%) without the unnecessary involvement of many resources or users (Nielsen, 2003)

With the aim to identify any difficulties in operating a digital library system features, Jones et al (2004) conducted an observational study to gather impressions of how people responded to the HistoryMap system. They found that although users found it easy to use the system, quickly identifying the meaning and purpose of the search location red circle, the timeline style of map search results, map navigation arrows and the overall browsing scheme, the more sophisticated features of the system were not used or fully understood by many of the participants as the features are not common on the Web (Jones et al., 2004). Measuring satisfaction and functionality of a system is the intention of most usability studies to find a way to articulate the usability of a specific digital library system and to recommend design changes that will create a more usable system as has been demonstrated in studies by Arko et al. (2006), France et al. (1999) and Phanouriou et al. (1999).

## METHODOLOGY

This study is primarily conducted to answer the following research question: How well does the developed prototype for the collaborative digital library perform in the management, creation, processing, searching and browsing of digital documents and objects in field trials in the digital library setting? The research question aims to evaluate the viability of a useful and enduring collaborative digital library for secondary school students.

In examining the needs of digital library stakeholders and how a collaborative digital library might be designed to meet these needs, Zachman Framework for Enterprise Architecture (Zachman, 2002) is used as the approach to investigate the user requirements and define the digital library organisation, processes, technology and information flows. The justification of using the framework, comparison with other existing digital library frameworks and mapping the artefacts and layers of Zachman framework to requirements analysis steps in building the digital library have been reported elsewhere (Abdullah, 2007a and 2007b; Abdullah and Zainab, 2004 and 2006). For an enterprise employment of the framework, Row Six of Zachman's represents the Functioning Enterprise, which is the end result of the architectural process (Zachman, 1987). In this research, the end result is to ensure that Row Six (the functioning system) represents what the stakeholders have in mind for the digital library enterprise. This paper reports on the assessment portion of Coredev. This involved
   (a) assessment of the usefulness of the system;
   (b) assessment of the usability of the system; and
   (c) site testing of the digital library.

A general user testing and evaluation procedure was conducted to sample users subjective view of the collaborative digital library prototype, on two aspects: usability and usefulness. An urban secondary school in the state of Selangor, Malaysia was chosen as the case sample. The testing and evaluation of the working prototype were conducted in two phases: (a) Task-based user setting, observations and questionnaire; and (b) user assessment via web-based survey questionnaire.

The first phase of user testing was conducted in three sessions, involving 12 Secondary Three (15 years old) students who were earlier a part of 30 students involved in a focus group interview who have volunteered to view and evaluate the digital library. These students had already completed and submitted their History project, and had indicated that they were willing to take part in the collaborative digital library project. All sessions were conducted at the digital library research laboratory at the Faculty of Computer Science and Information Technology, University of Malaya, with each session lasted for about 4 to 5 hours. Task-based user setting and direct observations are used to determine the overall amount of use and use of prototype's different features. First, the participants were given 15 minutes to explore the prototype. Then, they were specifically instructed to search and browse the digital library database, read the help instructions, and register as members of the collaborative digital library community. The 12 students were asked to:
   (a) Register as member, login and update their user profiles.
   (b) Browse the collections, create query specifications, use the simple or advance search and submit descriptive text information, examine the retrieved collections of search results, and display the contents of result items.
   (c) Upload a digital object and assign meta-labels to the digital object.
   (d) Create a report, assign meta-labels to the report and submit project report in the electronic format using the report generator tool. A sample project work (in the form of portfolio) is given to each participant as an example.

After the user testing session, participants were given a questionnaire, which was designed to elicit the participants' view regarding the usability and usefulness of the prototype. The questionnaire comprised 30 questions in three sections. The first section presents a Likert-scale type questions requiring a subjective satisfaction rating of Totally Agree, Agree, Somewhat Agree, Disagree, and Totally Disagree to 15 statements in relation to the usability and usefulness of the system. These 15 statements were adapted from the Software Usability Measurement Inventory (SUMI) (Kirakowski and Corbett, 1993; Macleod, 1994). SUMI is used because of the following reasons:
   (a) its validity and reliability have been established internationally;
   (b) it offers a convenient and inexpensive collection of trustworthy data;
   (c) only a minimum of about ten respondents is required;
   (d) it can be measured on a working prototype

The second section of the questionnaire presents a Likert-scale type questions requiring a subjective satisfaction rating of Totally Agree, Agree, Somewhat Agree, Disagree, and Totally Disagree to 13 statements in relation to the six system modules. The modules are (a) registration; (b) authentication; (c) data manipulation; (d) report submission; (e) search and retrieval; and (f) portal-enabled knowledge tool. The third section presents two open-ended questions to collect other user comments regarding technical problems during the field trials and suggestions on how the system can be improved.

The second phase of the user assessment was administered electronically. This was carried out on the improved prototype after the user feedback during the first phase of evaluation. The questionnaire was mounted on the collaborative digital library site (http://coredev.fsktm.um.edu.my) linked to a web-based survey tool (www.surveymonkey.com) for a period of time sufficient to gather at least 30 responses. Six History subject teachers who had agreed to collaborate in the projects were requested to inform and encourage their students to take part in the user assessment, after having used the digital library to create and submit their project work. Therefore, participation in the second phase was voluntary. The teachers were also given handouts about the collaborative digital library and instructions on how to use it to be distributed to their students who would voluntarily test the system.

A total of 44 users tested and evaluated the collaborative digital library; 12 students answered the lab questionnaire and 32 students took part in the web-based evaluation. The two phases of user assessment solicit students view on all functional requirements of the collaborative digital library except for the Indexing function in the Administrator's Module. Feedback on this Indexing module came from six postgraduate students registered in the course *WXGB6311 Digital Libraries* of the Masters of Library and Information Science (MLIS) Programme at the Faculty of Computer Science and Information Technology University of Malaya. The purpose of this usability testing is to predict the expected performance of the actual system administrators (teachers and teacher librarians) interacting with the current Indexing interface, as well as to detect any serious usability problems prior to the release of this service to teachers. The usability assessment of the Indexing Module evaluates the difficulties involved in using this function, as well as identifies possible future development work.

**FINDINGS**
This section presents the results obtained from the two phases of the user testing and evaluation, that is first, the laboratory questionnaire and second, the web-based questionnaire. The questionnaire required that users indicate the extent to which they agreed or disagreed with the 28 statements about the digital library according to a 5-point scale. The last two questions require an open-ended response which requested that users (a) describe any technical problems they had with the system; and (b) suggest a way to improve the digital library system so that it could better. An assumption is made that if a module (or function) scores 3.0 or above, it implies that the feature is well implemented.

**Motivated to Use: Users Feedback on Systems Overall Operation**
Motivation presents the users assessment of why they feel the collaborative digital library's overall operation is useful. Table 1 details the findings. Overall, users were satisfied with features of the collaborative digital library (Item 1, $\bar{x} = 3.75$). The users considered themselves capable of using the digital library easily (Item 3, $\bar{x} = 4.28$; Item 6, $\bar{x} = 1.98$; Item 12, $\bar{x} = 3.80$). In general, the overall feedback was positive since most users agreed that not only would they look forward to use the digital library for school project (Item 5, $\bar{x} = 4.05$), but would also consider recommending the system to friends (Item 10, $\bar{x} = 3.80$). The users found the interface very attractive (Item 4, $\bar{x} = 3.73$), a few students seemed favorably impressed with the interface finding it "simple", "wonderful" and "stimulating." One user indicated, "the interface is really good and people will be attracted to the good display of the information obtained in the digital library". A number of users expressed their appreciation for CoreDev with comments such as:
- "I think that the digital library is very good, personally I knew nothing about system and stuff before but with this [computer] programme, it has made me understand".
- "Using the system was much simpler and user-friendly than manually creating the report".
- "I don't know what other systems are like, but I would advise my friends to use this digital library".
- "The idea of a digital library for students is very good. Extend the idea to schools so that students can use it to do their scrapbook [projects]".

Table 1: Assessment of the Collaborative Digital Library Overall Operation (n = 44)

| Item No | Item Statement | Totally Disagree | Disagree | Somewhat Agree | Agree | Totally Agree | Mean |
|---|---|---|---|---|---|---|---|
| 1 | Overall I am satisfied with the CoreDev System | 2% (1) | 5% (2) | 18% (8) | 66% (29) | 9% (4) | **3.75** |
| 3 | I have learned to use the system with very little difficulty | 0% (0) | 2% (1) | 7% (3) | 55% (24) | 36% (16) | **4.28** |
| 4 | The interface of the system is very attractive | 2% (1) | 2% (1) | 25% (11) | 61% (27) | 9% (4) | **3.73** |
| 5 | I look forward to using this system at school | 0% (0) | 0% (0) | 16% (7) | 64% (28) | 21% (9) | **4.05** |
| 6 | Using this system is frustrating | 5% (2) | 93% (41) | 2% (1) | 0% (0) | 0% (0) | **1.98** |
| 10 | I will recommend this system to my friends | 0% (0) | 5% (2) | 13% (6) | 80% (35) | 2% (1) | **3.80** |
| 12 | I feel in command of this system when I am using it | 0% (0) | 5% (2) | 15% (7) | 75% (33) | 5% (2) | **3.80** |
| 13 | This system is really very awkward | 39% (17) | 50% (22) | 9% (4) | 2% (1) | 0% (0) | **1.75** |
| 15 | The system hasn't always done what I was expecting | 0% (0) | 63% (28) | 32% (14) | 5% (2) | 0% (0) | **2.41** |

Figure 1 presents the screenshot of the collaborative digital library main page. The menu on the left side of the main page consists of the following navigation buttons that intuitively describe the tasks these buttons should perform: Home, Login, Search, Browse, Game/Quiz, Feedback, About Us, FAQ, Help and Link. The terminology used in CoreDev interface was not a problem for the 'early adopters' who filled out the survey as no one commented on the English Language and the choice of lexicon used in the digital library. An important feature that enables users to see the dynamic side of the digital library is the introduction to "*Tokoh* of the Day" [Personality of the Day]. Two images are taken from the collection by using randomized technique, and as a result these images are alternately displayed each day on the main page of the digital library portal. "*Tokoh* of the Day'" will automatically display the image of the personality chosen for the particular day. The feature has the purpose of exposing the personalities to the users so that the users would be more aware of the many personalities and figures that play an important role. Besides, this helps to promote the content of CoreDev to the users and make them aware of the personalities that are being covered in the digital library.

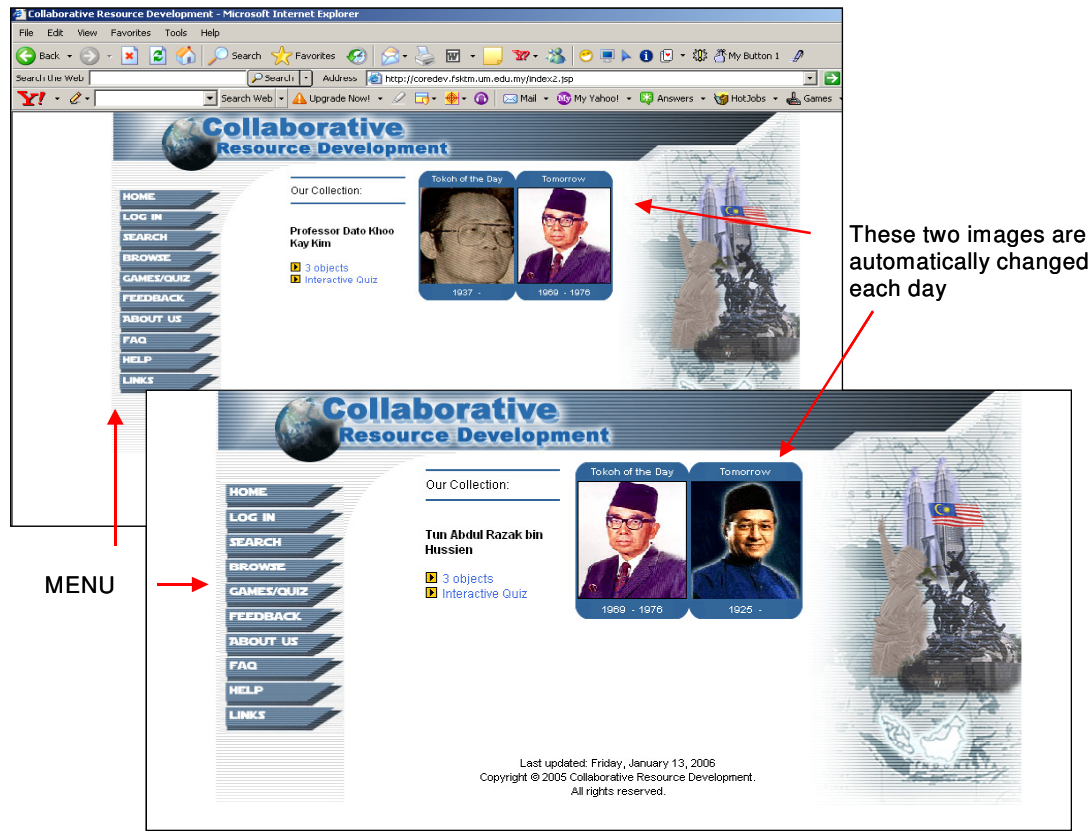

Figure 1: CoreDev Main Page Highlights "Tokoh of the Day" Displayed Using Randomized Technique

**Data: Users Feedback on the Ease of Handling Data**
The data dimension evaluates not only how the users handle the data, but also demonstrates the capabilities of the collaborative digital library functions to serve the communities as close to operational level as possible. Overall, users were satisfied with the functions or the programme modules of the collaborative digital library. Over 90% of the respondents indicated that they *Agree* or *Totally Agree* with the ease of use and comfort of five of the six features namely the registration module, authentication, data manipulation, report generator and search and retrieval (Table 2). However, the majority of the users (70%, 31) took a moderate stand on evaluating the portal enabled knowledge tool (Item 27, $\bar{x} = 2.77$). This tool is a value added function to the digital library where each day the portal would introduce or feature a specific collection. This function enables users to see the dynamic side of the portal since the main page of the portal would display a different collection everyday and make users be more aware of the various collections. A few users wanted the digital library to have more interesting and challenging educational games and quizzes. One commented that the games and quizzes are "too simple". This may be the reason why the portal enabled knowledge tool received the lowest assessment as at present it offers only three (3) personality quizzes consisting of 10 questions each, and two (2) games in the form of jigsaw puzzles. The following sub-sections present users perspective of how they handle data using the functions in CoreDev.

Table 2: User Assessment of the Collaborative Digital Library Programme Modules
(n = 44)

| Item No | Item Statement | Totally Disagree | Disagree | Somewhat Agree | Agree | Totally Agree | Mean |
|---|---|---|---|---|---|---|---|
| **Registration Module** | | | | | | | |
| 16 | I don't have any problem in signing up as new user | 0% (0) | 2% (1) | 2% (1) | 25%(11) | 70% (31) | **4.64** |
| 17 | The form to be filled in during registration is not too complex | 0% (0) | 5% (2) | 0% (0) | 20% (9) | 75% (33) | **4.66** |
| **Authentication Module** | | | | | | | |
| 18 | I can understand how to log in the system | 0% (0) | 0% (0) | 7% (3) | 27% (12) | 66% (29) | **4.59** |
| 19 | I know how to get my password if I can't remember it | 0% (0) | 5% (2) | 5%(2) | 31% (14) | 59% (26) | **4.45** |
| 20 | It is easy to log out the system | 0% (0) | 0% (0) | 2% (1) | 23% (10) | 75% (33) | **4.73** |
| **Data Manipulation** | | | | | | | |
| 21 | I know how to change my personal details after signing in | 0% (0) | 5% (2) | 5% (2) | 30% (13) | 61% (27) | **4.48** |
| 22 | It is not difficult to upload file for sharing | 0% (0) | 5% (2) | 18% (8) | 43%(19) | 34% (15) | **4.07** |
| **Report Submission** | | | | | | | |
| 23 | It is easy to create report using "Report Wizard" | 0% (0) | 2% (1) | 18% (8) | 68%(30) | 12% (5) | **3.89** |
| 24 | I feel comfortable reading the report produced by "report Wizard" | 0% (0) | 0% (0) | 25% (11) | 61%(27) | 14% (6) | **3.89** |
| **Search & Retrieval** | | | | | | | |
| 25 | The result of the search is accurate like I want | 0% (0) | 7% (3) | 25% (11) | 68%(30) | 0% (0) | **3.61** |
| 26 | It is easy to navigate the collection in the digital library | 0% (0) | 5% (2) | 11% (5) | 75%(33) | 9% (4) | **3.89** |
| **Portal-Enabled Knowledge Tool** | | | | | | | |
| 27 | Games and quizzes provided are suitable and interesting | 2% (1) | 23% (10) | 70% (31) | 5% (2) | 0% (0) | **2.77** |

**a) Registering and Managing User Information**

Users in general do not have problems signing in (Item 16, $\bar{x} = 4.64$) and logging out the system (Item 20, $\bar{x} = 4.73$). Testing of the registration form showed that it could be filled out quickly (Item 17, $\bar{x} = 4.66$). The users seemed fluent in managing their personal information (Item 19, $\bar{x} = 4.45$; Item 21, $\bar{x} = 4.48$). The Student Main Page, which brings users to tasks such as managing user profile and creating digital objects, appears upon successful login (Figure 2). For these purpose, users have the options of using two sets of menus: navigational buttons and animated buttons in the form of cubes. This page also provides statistical information related to the uploaded records of the user that logged in to the system. In this case, the user with Student ID 85, has a total of 7 uploaded records consisting of six (6) images and one (1) project report. The total size of records uploaded is also displayed; in this case the total size is 167.49 kilobytes.

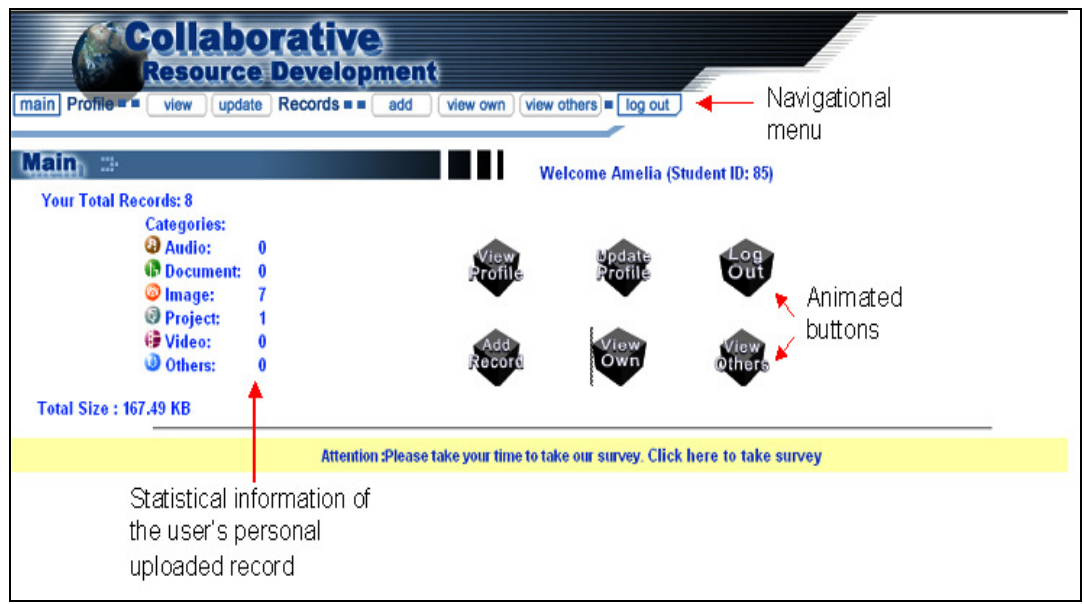

Figure 2: Student Main Page Upon Successful Login

**(b) Creating Report and Uploading Digital Objects**

User assessment indicates that users in general can use the authoring tools to create and submit resources into the digital library without difficulty. CoreDev supports two classes of authoring tools – for the novice and for the expert Internet user. The former is a report generator, which helps students create, organise and present their reports and it has the following features: a) a template to generate cover and background for the report; c) text editor that support organisation of heading and subheading, and various formatting features; d) uploading of images to be integrated within text; c) generation of appendices; d) generation of reference list according to the appropriate citation style; e) display and browsing of report. The latter supports uploading of reports and presentations, which may incorporate one or more types of multimedia contents. Using the upload file feature, students can upload files of various types into CoreDev to facilitate easy and organised retrieval and engage in information sharing. Both authoring tools support creating of description portion of the works by the contributors. The reference template in the Report Generator assists students to adhere to the correct citation style, record the digital sources used and the locations of those sources, to properly cite and credit those sources. This tool is needed as focus group interviews and document analysis of students' projects confirmed that students lack skills to accurately cite resources they use. Students select the type of resource they want to cite (either print or electronic), indicate the official standard they want to use, fill in the interactive form and the Wizard automatically formats the citation and display it in the students report.

Users agree that it is not difficult to upload digital objects into the digital library (Item 22, $\bar{x} = 4.07$). They also agree that it is easy to create report using the Report Generator (Item 23, $\bar{x} = 3.89$) and feel comfortable reading the report the generator produced (Item 24, $\bar{x} = 3.89$) (Table 2). Reactions to the report generator led to requests for more features in the text editor, more choices of report designs and background, users control of the number and location of images per page, and tutorial on how to describe the various portions of the report for quality metadata, as reflected in the following comments:

- "Please provide more templates for background and more tools for editing (font size, style, more font colours, resize tool for image, etc)".
- Allow user to create report background or upload images as background [for the report]
- "Report wizard is too restrictive. Students should have more choices to make report more creative and professional".
- "The location of image should be flexible, let user upload more than one image per page or per heading"
- "Need to include examples on how to fill up the textbox for description and keywords for quality information"

As a result from the user assessment of the report generator, the text editor has been improved to include more formatting tools such as more choices of font style and colours, as well as bullets and numbering. The number of templates for the background has been increased from five to ten, and users have more control to change the colour of the template. Users also are given the choices to align the image uploaded, either on the right, in the middle or on the left of the page. However the request to include more than one images per page as desired by the users will be handled in future enhancement.

**(c) Searching and Retrieving Information**
Overall, most participants were able to conduct searches, and find relevant items with little difficulty (Item 25, $\bar{x}$ = 3.61: Item 26, $\bar{x}$ = 3.89). There were no comments or suggestions made regarding the search and browse feature in the open-ended questions, although there were a few who disagree that the search result is accurate and the collections are easy to navigate (Table 2, item 24). CoreDev supports two types of search facilities, the simple search and the advanced search. The simple search is a Google-type box that basically provides free-text searching that will suit most new to experienced users, as the survey indicated that students in general underutilized advance search features of search engine. The advance search is a combination of two settings, which are the Type Delimiter setting and the Dropdown Menu setting (Figure 3). As illustrated, information seekers will be able to choose what best meets their needs based on these settings. In the Type Delimiter setting, each object type (such as documents, images, audio, video, hyperlinks and projects) can be unchecked to limit the search from retrieving the particular type. The Dropdown Menu provides the three available options: "Match Any Of These Words", "Match All Of These Words" and "Match Exactly This Phrase".

At the same time, a reasonable compromise between Google and a system to please an expert searcher who wants to search for specific occurrences of words is provided. The system has taken on this responsibility of assisting the users to fine-tune their queries based on attributes such as creator, keywords, collection and resource type through multi-criteria search settings, Students can learn the more fluent use of search tools, mainly in the capacity to narrow and revise searches to better specify what they want. The search feature also enables the users to save certain search preferences to make the searching process more effective and efficient. Four (4) preferences settings can be performed to suit individual likings. These preferences settings are Interface Language (English or Malay Language), Query Box Size, Number of Results and Results Window (Figure 4). If the user's goal is browsing, s/he may view the resources by collection, period (year), resource type, alphabetical order, and thumbnail images. The Browse Audio, Browse Video, Browse Hyperlinks and the Browse Projects pages share much of the same interface and functionality as the Browse Documents page Browsing is based on Modified Dublin Core metadata and the historical collections are also categorized based

*Abdullah, A & Zainab, A.N*

on multimedia type to facilitate users to choose based on categories. The items returned that match the query parameters can be evaluated by their textual description, thumbnail and browse images, and metadata attributes.

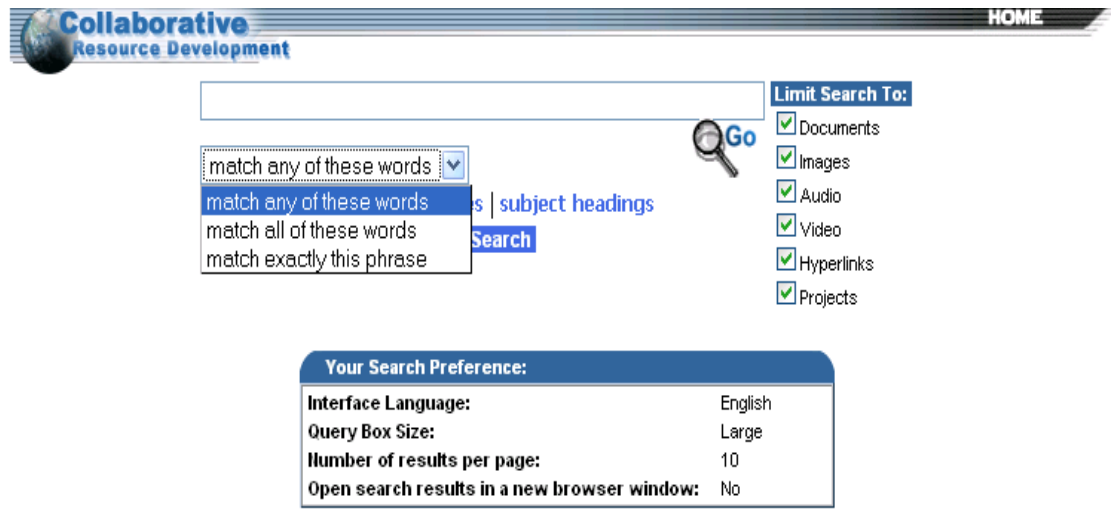

Figure 3: Simple Search Combining Two Settings – Type Delimiter and the Dropdown Menu

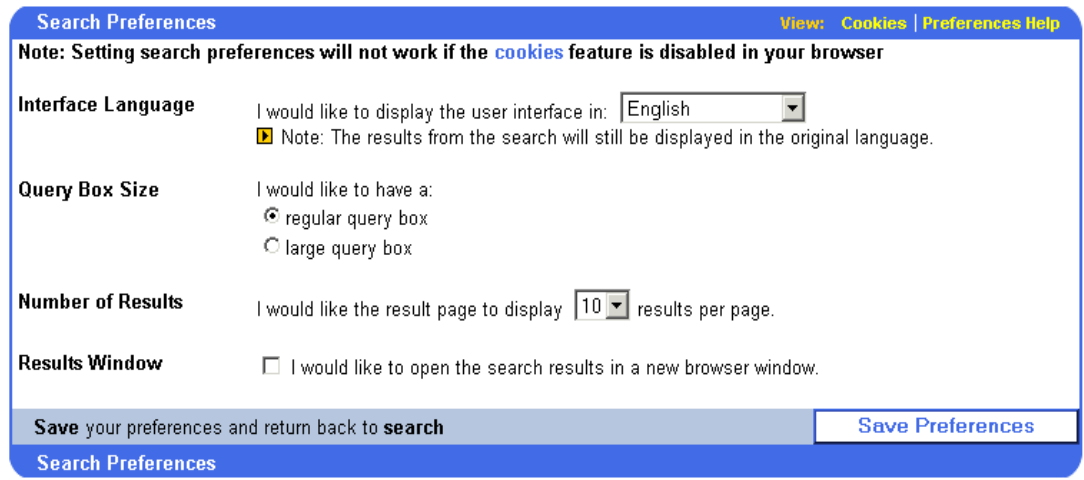

Figure 4: The Search Preference Page

**(d) Indexing the Digital Library Resources**
An important feature to facilitate resource discovery of digital objects in CoreDev is the Indexing module. Before any object can be searched or made accessible to the users, it must first be indexed using the indexing function, which is incorporated into the Administrator's module. In this module, an administrator (teacher or teacher librarian) performs six exclusive tasks namely grading the project reports, editing the collection, indexing digital objects, editing entries, defining object to objects relations and assigning new subject headings. The administrators can extract information about the registered users and also produce statistical reports whenever required using the Tracking and Reports feature. The administrator is also able to refine or fine-tune the ranking values for the relevancy ranker. There is also a preference setting whereby the administrator can set the number of records to be displayed in each administrative page and the option to make changes to his or her particulars.

As this module is used by teachers or teacher librarians, the questionnaire did not solicit users (who registered as student) opinion on the indexing module. The user evaluation of the indexing modules by the six MLIS students indicates that teachers need to be trained in publishing digital resources and in indexing as well as validating the resources to ensure that the digital library contents can be efficiently retrieved, as even the MLIS students, who in general are ICT literate, have difficulties with the module. Reactions to the user interfaces of Index Upload Materials, Indexing Templates, as well as Edit Entries, led to requests for better tutorials and context-sensitive help, other terminology or words used to define tasks, viewing objects feature to define relationships, and better quality control of the metadata as reflected by the following verbatim comments:
- "There is insufficient HELP to explain what each function does, for example index and edit, and where the objects are after uploading."
- "No clear description as to what each field is for in the Indexing Module."
- "Very confused between Index function, Relationship function. Also, difficult to relate 'Collection' in Index function and how to assign 'Collection'. Change word 'Index' to 'Classify' or 'Catalogue'."
- "Should include an example in every metadata fields so that new indexers will be able to learn how to input the descriptors.
- "There should be examples of how to fill a metadata form by the side of the input box in administration login in order for us to understand what to put in the input text box or the upload file location input"
- "I face the problem of creating the relation in between object since I have to remember the ID by myself."

- "Relationship between objects is difficult to ascertain unless the indexers could view the file or footages while indexing".
- "Some of the information required is not easy to input."

**(e) Expanding the Collection**
In the open-ended question, most of the users expressed support for expanding the database, for example "I hope you are continuing to add more resources – the concept of digital library for students project is excellent" and "I want to see more topics, more *Tokoh* (personality)". Some respondents made specific suggestions for additions, such as "more games and quizzes", "more resources on "*Peristiwa Bersejarah*" (historical events) to balance what is already in place", "more collections on *Tokoh* (personality)", and "more local history, local historical events". One user specifically suggested the school to make available the current History projects in the electronic form as indicated by the following comment: "Softcopy the current History project [reports]". Some even recommended specific resources to add to the site for example history tests, question banks and teachers' notes. This may seem that the lack of extensive content in CoreDev can be a problem for users trying to make the system work for them. The immediate implications from these findings are that CoreDev needs to put a high priority on increasing the contents of the system and on providing a redesigned web interface to access those contents and to inform users about the limitations of content. At present, CoreDev has developed a total collection of 777 resources consisting of consisting of 126 documents, 35 projects, 437 images, 23 audios, 34 videos and 90 hyperlinks obtained from both the report generator and the upload function (Table 3).

Table 3: Digital Resources of Various Types

| Object Type | Via report generator | | Via upload object function | | Total number of digital objects | |
|---|---|---|---|---|---|---|
| | Count | Percentage | Count | Percentage | Count | Percentage |
| Audio | 17 | 3.6 | 6 | 2.0 | 23 | 3.0 |
| Document | 103 | 21.8 | 23 | 7.6 | 126 | 16.2 |
| Hyperlink | 90 | 19.0 | 0 | 0.0 | 90 | 11.6 |
| Image | 213 | 45.0 | 224 | 73.7 | 437 | 56.2 |
| Project | 26 | 5.5 | 9 | 3.0 | 35 | 4.5 |
| Video | 24 | 5.1 | 10 | 3.3 | 34 | 4.4 |
| Others | 0 | 0.0 | 32 | 10.5 | 32 | 4.1 |
| Total | 473 | 100.00 | 304 | 100.0 | 777 | 100.00 |

*Collaborative Digital Library of Historical Resources*

**People: The Digital Library Participant Description**
People describes the registered users of the collaborative digital library who should comprise the students and teachers. The 44 users who completed the survey constitute approximately 42% of the beta testers who actually used the system at least once. The researcher is handicapped by drawing any inferences from this data as the survey population lacked sufficient representation of user characteristics such as Internet usage and proficiency. There were 105 beta testers comprising 59% (62) girls and 41% boys (43) registered as students. The student registration data obtained from the Report and Tracking Menu in the Administrator Module was analyzed to see what could be learned about the interested users who wanted to test CoreDev. Based on the number of digital objects created as "project", it appears that about 45% of the beta testers actually used CoreDev to create and submit complete History project reports (totaled 20 out of 99), whereas others merely tested the report creation tools. There are five (5) schools in Selangor State and one (1) school in Federal Territory of Putrajaya represented, with 73.3% (77) of the users coming from the case school since the students from this school participated in the study and were already informed about the digital library. Secondary Three students comprise the majority of the users (50.5%, 53), followed by Secondary Two students (14.3%, 15). Figure 5 presents the participants by gender and secondary level. The findings suggest that that the collaborative digital library is reaching its target educational communities and there is a possibility that information about CoreDev is being disseminated to students from other schools, and students from the upper secondary level (Secondary 4 and 5).

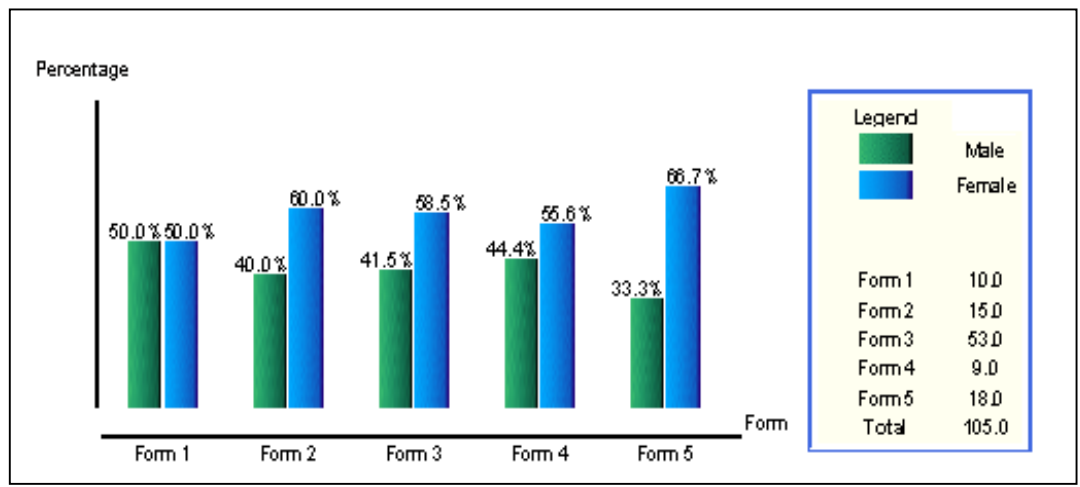

Figure 5: The Collaborative Digital Library Users by Gender and Secondary Level (n = 105)

*Abdullah, A & Zainab, A.N*

**Function: Users Feedback on Procedural and System Documentation**

The Process dimension presents users feedback on the procedural and system documentation of the collaborative digital library. Analysis on 5 item statements from the Likert-scale questionnaire indicates that the users liked the tutorial and the ease of exploring the system (Table 4). Users understand how the digital library functions (Item 9, $\bar{x} = 3.82$). They feel the instructions and prompts are helpful (Item 7, $\bar{x} = 3.89$) and many of them disagree that they do not know what to do next when navigating the system (Item 14, $\bar{x} = 1.98$). The reason for these positive feedback may be attributed to the provision of messages after the user has completed a particular task such as successful registration, uploads and submission of reports and signing out of the system. These messages are helpful and important for the users because it allows the users to be kept up-to-date with what is going on and informed of what has been done to the system. Users also agreed that help information on how to use the system is sufficient (Item 28, $\bar{x} = 3.70$). They disagree that they have to look for assistance most of the times (Item 11, $\bar{x} = 2.11$).

Table 4: Users Feedback on Procedural and System Documentation (n = 44)

| Item No | Item Statement | Totally Disagree | Disagree | Somewhat Agree | Agree | Totally Agree | Mean |
|---|---|---|---|---|---|---|---|
| 7 | The instructions and prompts are helpful | 0% (0) | 2% (1) | 18% (8) | 68% (30) | 12% (5) | **3.89** |
| 9 | I understand how the digital library system functions | 0% (0) | 2% (1) | 18% (8) | 75% (33) | 5% (2) | **3.82** |
| 11 | I have to look for assistance most of the times when I use this system | 9% (4) | 75% (33) | 12% (5) | 5% (2) | 0% (0) | **2.11** |
| 14 | I sometimes don't know what to do next with the system | 18% (8) | 68% (30) | 12% (5) | 2% (1) | 0% (0) | **1.98** |
| 28 | Help information on how to use the system is enough | 0% (0) | 7% (3) | 20% (9) | 68% (30) | 5% (2) | **3.70** |

**Network: Users Feedback on the Robustness of the Network Using Chosen Communication Facilities**

The Network dimension describes the users' feedback on robustness of the network in handling data from various locations using chosen communication facilities. CoreDev presence is currently manifested as a web portal at http://coredev.fsktm.um.edu.my. This was made known to the Secondary 2 and 3 students who were encouraged to test CoreDev through creation and submission of their History project. User assessment indicated that CoreDev is robust enough in handling data due to its response time. A total

116

of 27 users felt that the system speed is fast enough (62%), although three (3) users found that the system was a bit slow (Table 5). This may be due to several reasons, including the slow response of the user's PC, network lines, server and peak timings. None of these users however indicated specific problems regarding network connection when they accessed and tested the digital library system. This reflects that there is a reliable and active line connection between the points (users) set up by a telecommunication common carrier and indicates that CoreDev does not exhibit serious drawbacks in terms of speed.

Analysis of the data on 105 beta testers indicates that users use the system and upload digital resources from various locations in Petaling Jaya and Shah Alam, Selangor, as well as from Putrajaya as revealed from the findings on the network location of the digital library participants, reflecting that the system is able to handle uploads of digital objects from various locations. The system is also able to handle uploads of digital objects of various format and file size based on the data analysis on digital objects uploaded from 1 March 2005 to 15 September 2006, which indicates that a total of 304 files of various formats were uploaded with a minimum file size of 0.81 Kb and a maximum of 2 664Mb ($\bar{x}$ = 223 Mb).

Table 5: User Assessment on the Response Time of the Collaborative Digital Library
(n = 44)

| Item No | Item Statement | Totally Disagree | Disagree | Somewhat Agree | Agree | Totally Agree | Mean |
|---|---|---|---|---|---|---|---|
| 8 | The speed of this system is fast enough | 2% (1) | 5% (2) | 32% (14) | **57% (25)** | 5% (2) | **3.57** |

**Time: Users Feedback on Systems Operation Schedule**
Time describes users' feedback on CoreDev system operation schedule related to time. It specifically determines when users use the system and seek users' feedback on the time they take to use the system. Data from the Report Tracking Module, which reports the date and time users upload resources onto CoreDev shows that the system is in operation 24/7. Out of 304 uploads from 1 March 2005 to 15 September 2006, a total of 65.5% occurred during the day (7.00 a.m. to 7.00 p.m.) with the mean time of 1.00 p.m.

Table 6 indicates that the majority of the users (82%, 36) agree with the statement that it has taken very little time to use the system (Item 2, $\bar{x}$ = 4.02). However, it was observed during the first phase of user assessment that, in creating and submitting project reports, not all the users showed fluency to deal with the prototype (it took them from twenty to thirty minutes to complete the task), although they are experienced users in terms of e-content creation. This maybe because the tasks involved more actions such as formatting, embedding images, describing portions of work and compiling references, than merely typing and submitting documents.

Table 6: User Assessment on the Time taken to Use the Collaborative Digital Library
(n = 44)

| Item No | | Totally Disagree | Disagree | Somewhat Agree | Agree | Totally Agree | Mean |
|---|---|---|---|---|---|---|---|
| 2 | It has taken very little time to use the system | 0% (0) | 0% (0) | 18% (8) | **61% (27)** | 21% (9) | **4.02** |

Although the majority of the users agree that the speed of the system is fast enough as reported earlier, many of the comments in the open-ended question expressed problems or difficulties associated with time. Most of the technical problems reported in the survey were associated with the response time where it took users some time to log in and to upload a file. In those situations, consistent with findings by Nielsen (2000) and Ferreira and Pithan (2005), the users' most common feelings were discomfort, impatience and frustration besides the great deal of time spent to finish the task. A number of users expressed their difficulties using the system with comments indicated below:
- "Station time-out should be extended and [system] should alert [users] when it logs out on its own
- "The log in speed is a bit too slow, is like hanging there for few minute, it might confuse other, that the system is not running or having problem".
- "Log in [is] a bit slow, some time uploading a file takes too long to complete"

**Other Suggestions**
The survey solicits users' feedback on how to improve the performance of the digital library. There were two categories of recommendations, which the researcher rated as high priority, and CoreDev should be able to support these in future. First, there was a consistent request for a way for the user to address another person. Users suggested more support should be provided apart from an email address, Feedback, FAQs, and static Help pages. Users would like to have discussion board to generate topics and problems and communicate regarding tasks and information problems, either with teachers, or among group of students. This calls for more collaborative features that require mechanisms for users with common interests to identify one another as well as tools for conducting the discussion.

The second suggestion was that users would like to write comments, or view comments or feedback on the project reports they view. This requires students and teachers to be

*Collaborative Digital Library of Historical Resources*

able to annotate directly on the digital report. Document analysis of students' project indicates the need for an annotation tool where incorrect information can be highlighted or commented by teachers, so that the use of incorrect information does not perpetuate in future reports (Table 7). However, the need for this tool did not surface during the survey and interviews. This would be a feature for future enhancement. Table 7 details the users' suggestions leading to the need for CoreDev to have support for communication and annotation.

Table 7: Expressed Need for Communication and Annotation Support

| **On the need to have a communication tool….** | **On the need to have an annotation tool….** |
|---|---|
| [Can we chat while doing our report] <br> More support such as discussion group <br> [Anyone can pose about a problem, anyone can answer] <br> Comments and feedback not enough, more support to communicate to system administrator <br> Can we have a toll free 1-800 to ask about this system when we have problem <br> If I have a problem I want to ask other people who have used this digital library. Can you have that? <br> Should have discussion or bulletin board | [We can comment about other people's report] <br> Can see and comment friends work when they are preparing (report) I mean when they have not submitted to teacher <br> [Maybe let users rate people's report or the information they upload if we find the information useful] <br> I like to comment on the reports or information people upload <br> [Let teachers comment our work first before we submit] |

Note: Responses in [ ] denote translation from Malay Language

**CONCLUSION**

This paper has presented user assessment of the digital library prototype to gauge the viability of a useful and enduring collaborative digital library for school projects. The testing and evaluation was conducted to gather feedback on the satisfaction level, technical problems and suggestions for improvement to the collaborative digital library prototype. In summary, the feedback from the first users are as follows: (a) Overall, users were satisfied with features of the collaborative digital library. (b) Most participants were able to conduct searches and browsing, and find relevant items; (c) Users felt that it was easy to create reports using the report generator or the uploading module; (d) Users' general impression about the digital library was that it is a pleasant site in terms of visual aspects, organisation and distribution of information; (e) Users reported that it has taken them very little time to learn and use the system and they look forward to using the system at school; and (e) Users demonstrated easiness in learning (learnability) and remembering the steps they had taken to perform the task (memorability), when asked during the post-test interview. Also, users' feedback pointed out the need to support the collaborative digital library with a communication and annotation tool. The potential to add value to usage of digital resources by using various approaches to synchronous and asynchronous communication as well as annotation tools is illustrated by the large number and diverse approaches and implementations of such systems on the Web.

As a testbed system, the collaborative digital library exists more to demonstrate capabilities than to serve communities as has been reflected by the positive feedback on the functional requirements as compared to the feedback on content. The beta tester demographics and user survey results indicate that the collaborative digital library is reaching its target communities, and can potentially tell how satisfied those communities are with CoreDev. A number of useful comments and suggestions were put forward in the open-ended question. These responses were the most revealing outcome of the evaluation exercise as the statements were specific and insightful. However, the comments, especially on the technical problems users had with the system, painted a somewhat more negative picture of the digital library prototype than did the Likert-scale responses. This is consistent with the idea that people are motivated to comment when they encounter problems (Hill et al, 2000). The problems are associated mainly with file size and response time. Although the users were faced with technical difficulties during the accomplishment of the tasks, observations and interview indicated that they felt satisfied upon completing the task, in which they were presented with digital library services and resources not encountered before or other forms of search task performance that they did not know. They were also pleased to see their work published in the digital library. The results imply that such a digital library could be usefully utilised to support project-based learning and subsequently inculcate ICT and information literacy amongst Malaysian educational communities.